\documentclass[aps,prl,twocolumn,showpacs]{revtex4-1}
\usepackage{graphicx}
\usepackage{amssymb}
\usepackage{float}
\begin{document}
\title
{\bf Weakly Anisotropic Noncentrosymmetric Superconductors with Radial Line Nodes and Thermodynamical Anomalies}
\author{Mehmet G\"{u}nay$^{\bf (1)}$, T. Hakio\u{g}lu$^{\bf (2,3)}$ and Hasan H\"{u}seyin S\"{o}mek$^{\bf (3)}$}
\affiliation{ 
${\bf (1)}$ {Department of Physics, Bilkent University, 06800 Ankara, Turkey}
\break
${\bf (2)}$ {Consortium of Quantum Technologies in Energy (Q-TECH), Energy Institute, \.{I}stanbul Technical University, 34469, \.{I}stanbul, Turkey}
\break
${\bf (3)}$ {Department of Physics, \.{I}stanbul Technical University, 34469, \.{I}stanbul, Turkey}
}
\begin{abstract}
In noncentrosymmetric superconductors (NCSs), the inversion symmetry (IS) is most commonly broken by an antisymmetric spin-orbit coupling (SOC) removing  the spin degeneracy and splitting the Fermi surface (FS) into two branches. A two component condensate is then produced with a doublet pair potential mixing an even singlet and an odd triplet. When the triplet and the singlet strengths are comparable, the pair potential can have rich nodes. The angular line nodes (ALNs) are associated with strong anisotropy and they are widely studied in the literature. When the anisotropy is not strong, they can be replaced by other types of nodes in closed or open forms affecting the low temperature properties. 

Here, we focus on the weakly anisotropic case and the line nodes in the superconducting plane which become circular in the limit of full isotropy. We study the topology of these radial line nodes (RLNs) and show that it is characterized by the $Z_2$ classification similar to the Quantum-Spin-Hall Insulators. From the thermodynamical perspective, the RLNs cause, even in the topological phases, an exponentially suppressed low temperature behaviour which can be mistaken by nodeless s-wave pairing, thus, providing an explanation to a number of recent experiments with contraversial pairing symmetries. In the rare case when the RLN is on the Fermi surface, the exponential suppression is replaced by a linear temperature dependence. The RLNs are difficult to detect, and for this reason, they may have escaped experimental attention. We demonstrate that Andreev conductance measurements with clean interfaces can efficiently probe the weakly anisotropic samples where the RLNs are expected to be found. 
\end{abstract}
\maketitle

\section{1-Introduction}
Superconducting symmetries beyond the conventional BCS spin singlet pairing were known since 1960s. Distinct examples are $^3He$ \cite{He3}, heavy fermion \cite{Heavy_Fermion}, high $T_c$ \cite{High_Tc} superconductors as well as the NCSs\cite{NCS_review,NCS_review2}. Strongly momentum dependent electronic correlations, broken spin-degeneracy, broken IS and the SOC in these systems, add to the variety of factors leading to exotic spin and momentum dependent phenomena together with the unconventional pair formation\cite{TH,OPs,dz}. The manifested time reversal symmetry (TRS), its spontaneously broken (TRSB)\cite{TRSB} phases and the non trivial topological properties of the  electronic bands add to the plethora that make understanding of these unconventional effects an experimental and theoretical challenge\cite{USC}. 

The NCSs are classified as unconventional superconductors with broken IS while a substantial part of superconductors has this symmetry. The IS is most commonly broken by the presence of a SOC. Let us confine our attention to the two dimensions, i.e. ${\bf k}=(k_x,k_y)=k\,(\cos\phi,\sin\phi)$ as the wavevector of the superconducting electrons in the $x-y$ plane. Since the parity and the spin are not conserved, i.e. the OPs can simultaneously accommodate an even singlet $\psi_{\bf k}$ and an odd triplet ${\bf d}_{\bf k}=(d_{x\bf k},d_{y\bf k},d_{z\bf k})$. The full $2\times 2$ pair potential in the spinor basis is then given by $\Delta({\bf k})=i(\psi_{\bf k}+{\bf d}_{\bf k}.{\bf \sigma})\sigma_y$ and the connection is made with the singlet and triplet OPs by $\psi_{\bf k}=[\Delta_{\uparrow \downarrow}({\bf k})-[\Delta_{\downarrow \uparrow}({\bf k})]/2$, $d_{x,{\bf k}}=[\Delta_{\downarrow \downarrow}({\bf k})-[\Delta_{\uparrow \uparrow}({\bf k})]/2$, $d_{y,{\bf k}}=[\Delta_{\uparrow \uparrow}({\bf k})+[\Delta_{\downarrow \downarrow}({\bf k})]/(2i)$ and $d_{z,{\bf k}}=[\Delta_{\uparrow \downarrow}({\bf k})+[\Delta_{\downarrow \uparrow}({\bf k})]/2$. Here, we choose the spin quantization axis perpendicular to the $x-y$ plane. The roles played by the TRS and the Pauli exchange symmetries are crucial. The former enforces $\Delta_{\nu \nu^\prime}({\bf k})=\beta\,\Delta_{{\bar \nu} {\bar \nu}^\prime}^*(-{\bf k})$ where ${\bar \nu}$ is the opposite spin to $\nu$ and $\beta=+1$ for $\nu=\nu^\prime$ and $\beta=-1$ for $\nu \ne \nu^\prime$. The Pauli exchange requires $\Delta_{\nu \nu^\prime}({\bf k})=-\Delta_{\nu^\prime \nu}(-{\bf k})$ under the exchange of paired fermions. When these symmetries are both manifested, an important result follows that $\psi_{\bf k}$ and ${\bf d}_{\bf k}$ are both real. 

The smoking gun of the unconventional pairing in a NCS is usually considered as the nodes of the pair potential\cite{NCS_review,evidence1,evidence2,early_nodes}. The nodal structure can most simply be represented in two dimensions in the angular momentum-$L_z$ basis as $X_{\bf k}=\sum_m \,X^{(m)}_{\bf k}$ where $X^{(m)}_{\bf k}={\cal Y}_m({\hat {\bf k}})X^{(m)}_k$, ${\hat {\bf k}}={\bf k}/k$ and ${\cal Y}_m({\hat {\bf k}})\propto (\cos{m\phi}, \sin m\phi)$ are angular momentum basis functions describing the anisotropy\cite{Sigrist_Ueda} with $X_{\bf k}=(\psi_{\bf k},{\bf d}_{\bf k})$. Here $X^{(m)}_k=(\psi^{(m)}_k,{\bf d}^{(m)}_k)$ are defined as the OPs for each partial component $m$ which will be elaborated below. 

In two dimensions as we consider here, we represent the IS breaking by the SOC vector ${\bf G}_{\bf k}=\alpha (-k_y,k_x,0)$ with $\alpha$ as the SOC constant. The nodes in the pair potential are represented by a) discrete set of points, i.e. point nodes (PNs), and b) closed or open line nodes in the ${\bf k}$ space. The excitation spectra are symmetric around all TRS-invariant points such as ${\bf k}=0$ and ${\bf k}=\{(\pm \pi/a,0), (\pm 0,\pi/a)\}$ where $a$ is some lattice parameter. Note that, in strongly anisotropic compounds with the tetragonal symmetry $C_{4v}$, these are special points supported by the center and the boundaries of the Brillouin zone. Furthermore, under manifested TRS, the excitation spectra are Kramers degenerate $E_{\bf k}^\lambda=E_{-\bf k}^{\lambda^\prime}$ with $\lambda, \lambda^\prime$ describing different branches split by the broken IS. PNs can occur at these TRS-invariant points or, in the case of Weyl points they may occur at arbitrary values in a finite number of TRS  pairs\cite{Weyl}.  

In addition to the PNs in strongly anisotropic two dimensional NCSs, a common type is the angular line node (ALN). In tetragonal systems, the nodes can be formed along $k_x=\pm k_y$ or ${\bf k}=(0,k_y)$ or $(k_x,0)$. In low temperatures, ALNs are evidenced by integer scaling exponents in the specific heat and this has been observed in a number of highly anisotropic NCSs\cite{evidence1,evidence2} among which are the celebrated TRS preserving $CePt_3Si$ and the TRS breaking $Sr_2RuO_4$. In the presence of ALNs, the London penetration depth, heat conductivity and ultrasound attenuation all scale with integer exponents. However other experiments also exist where ALNs cannot sufficiently explain the thermodynamic data\cite{Sigrist_Ueda} and, despite this large number of experimental and theoretical studies, a one-to-one connection between the scaling behaviour and the existence of nodes is missing. It is remarkable that the anisotropy, strong electronic correlations and the nodes are common topics that are usually discussed together\cite{NCS_review2}. %

For the spin singlet-triplet mixed NCSs with a specific point group symmetry, let us consider the leading terms in the angular momentum expansion. For the tetragonal case these terms are the $s$ ($m=0$) and the $d$-wave ($m=2$) components in the singlet $\psi_{\bf k} \simeq \psi_k^{(0)}+\psi_k^{(2)} \cos 2\phi$ and the $p$ and the $f$-wave components of the triplet ${\bf d}_{\bf k}\simeq {\hat {\bf k}}\,(d_k^{(0)}+d_k^{(2)} \cos 2\phi)$. The mixed state pairing potential $\Delta_{\bf k}^\lambda=\psi_{\bf k}-\lambda \gamma_k F_{\bf k}$, where $\lambda=\pm$ is the band splitting due to the broken IS, is directly responsible for opening an energy gap at the Fermi level as well as giving rise to a topological band structure. Here, $F_{\bf k}=\vert {\bf d}_{\bf k}\vert$ and $\gamma_k$ is a function of $k$ which can only take the values $\pm 1$ (see Section.2). In our case, the $\Delta_{\bf k}^\lambda$ is given by   
\begin{eqnarray}
\Delta_{\bf k}^\lambda=A_k^\lambda+B_k^\lambda \cos 2\phi
\label{anisotrop-isotrop_pairing}   
\end{eqnarray}
 where $A_k^\lambda=\psi_k^{(0)}-\lambda \gamma_k F_k^{(0)}, B_k^\lambda=\psi_k^{(2)}-\lambda \gamma_k F_k^{(2)}$ are $m=0,2$ components respectively. 
The ALNs are obtained in the strongly anisotropic case of $\vert A_k^\lambda/B_k^\lambda \vert \le 1$ in the small wavelength limit. Specifically, this is the regime where the effect of the short range electronic correlations in the form of SDWs, CDWs or Fermi surface nesting become important. On the other hand, in the other extreme case of weak anisotropy the nodes are dominated by the first term in Eq.(\ref{anisotrop-isotrop_pairing}) where they are closed lines encircling the origin. 

\begin{figure} 
\includegraphics [scale=0.2]{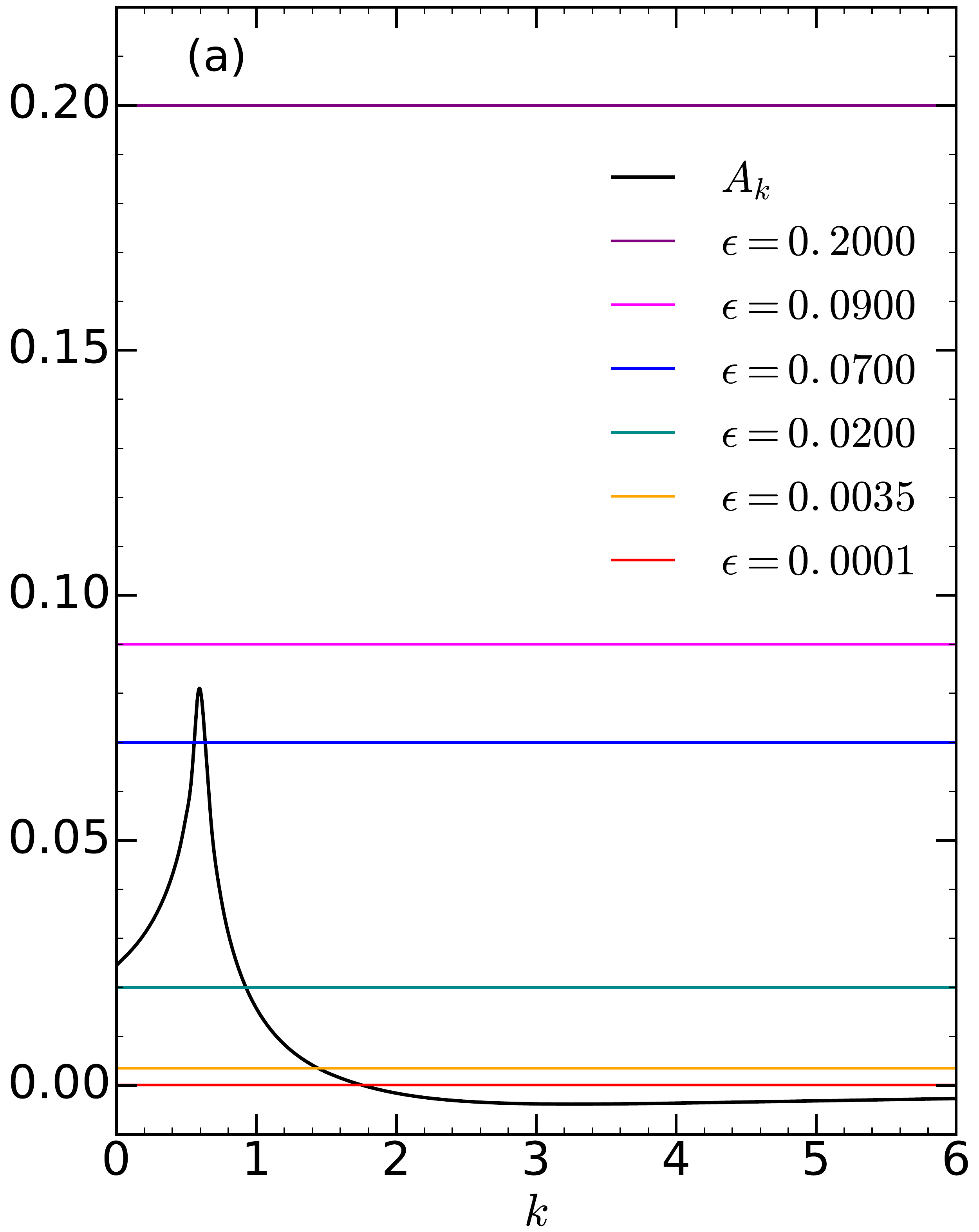} \\
\includegraphics [scale=0.28]{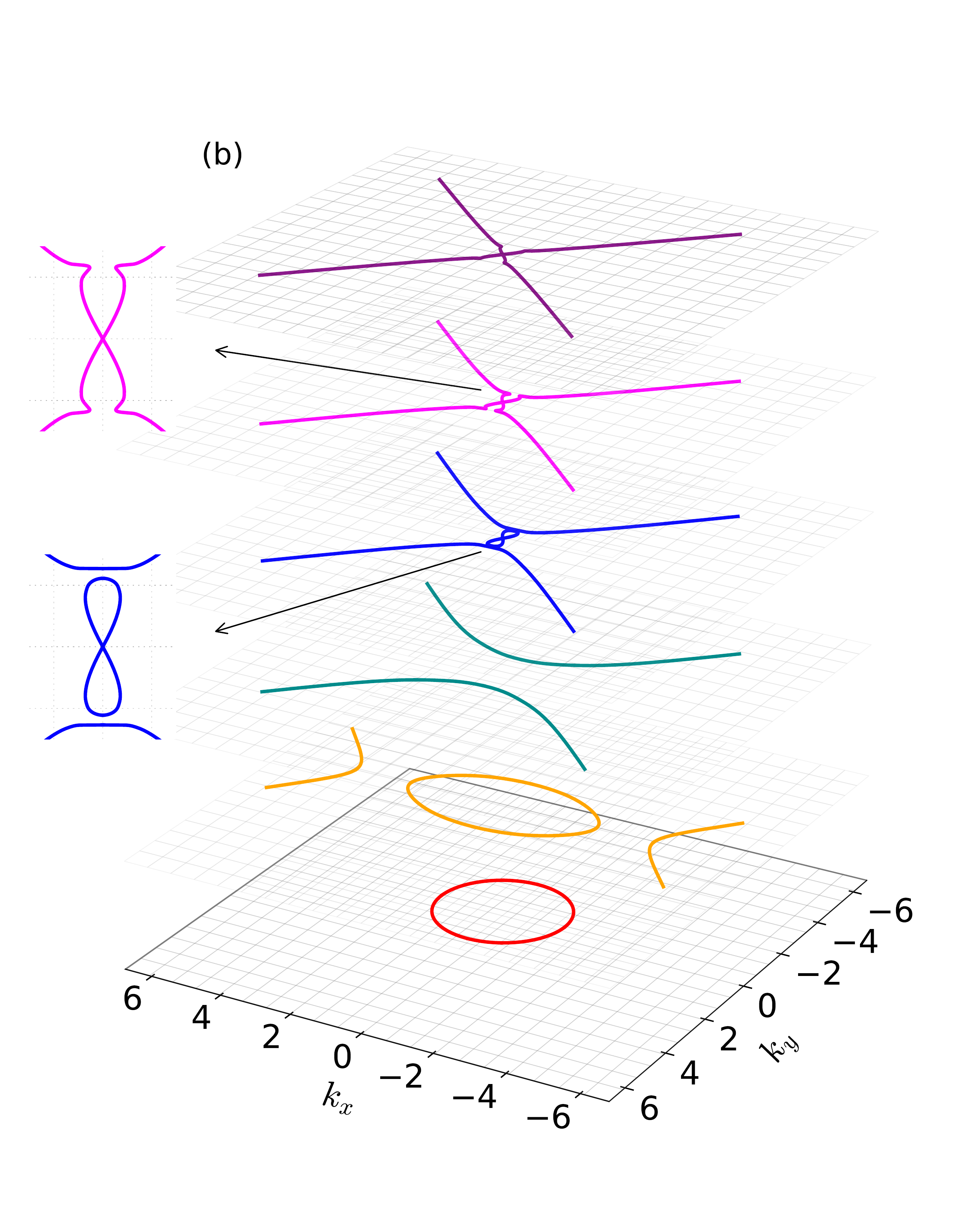}
\caption{(Color online) The transmutation of the nodes of $\Delta_{\bf k}^\lambda$ in Eq.(\ref{anisotrop-isotrop_pairing}) for a fixed $\lambda$ and for $A_k^\lambda$ as shown in (a). Different values of $B_k=\epsilon$ from a strongly anisotropic (upper most for $\epsilon=0.2$) to the weakly anisotropic case (lower most for $\epsilon=10^{-3}$) are indicated on the vertical scale in (a) and the corresponding nodes of $\Delta_{\bf k}^\lambda=0$ are shown in (b) with the same color coding.}  
\label{ALN_to_RLN}
\end{figure}

The extreme limits of anisotropy is therefore determined by the leading term being $A_k$ or $B_k$ in Eq.(\ref{anisotrop-isotrop_pairing}). Since the magnitude of the second term is limited by $\vert B_k \vert$, we use a simple model where $B_k=\epsilon$ is constant and $A_k$ is given by the profile in Fig.\ref{ALN_to_RLN}.(a). Using this $A_k$ in Eq.(\ref{anisotrop-isotrop_pairing}), the change from the fully isotropic RLNs to the strongly anisotropic ALNs can be studied as $\epsilon$ is changed. This transmutation changes the rotational symmetry of the energy gap. The Eq.(\ref{anisotrop-isotrop_pairing}) permits a number of other intermediate solutions including nodal arcs and point nodes. These unconventional forms originate from the fully self-consistent solutions starting from the anisotropic pairing interaction and the renormalized electronic bands. Revealing this nodal variety by experimental techniques is also important. In this context, a {\it superconducting node engineering} may be developed in the future which can provide valuable information about the elusive mechanisms behind the unconventional pairing and, not the least new and exotic device applications. An example of the intermediate regime between RLNs and ALNs is shown in Fig.\ref{anisotrop-isotrop_pairing}.(b).    

It is commonly stated in the literature that, the weak anisotropy is a signature of the lack of strong correlations leading to the s-wave pairing and trivial topology. In this report it is shown that weakly anisotropic NCSs can have non-s wave pairing with nontrivial topology if RLNs are present. This important point is where our motivation resides in this work. We believe that the results of our work are relevant to some of the experiments where the effect of the lattice symmetry on the electronic system is weak. With the IS broken, the relevant electronic symmetry can then be approximated by $C_{\infty v}$ which corresponds to the circular symmetry in the lowest slice of the Fig.\ref{ALN_to_RLN}.(b). In this case, the $m=0$ component dominates with $\vert\psi_{\bf k}\vert \simeq \psi_{k}^{(0)}$ and $\vert{\bf d}_{\bf k}\vert \simeq F_{k}^{(0)}$ and we simply drop the $(0)$ index. In the pair potential in Eq.(\ref{anisotrop-isotrop_pairing}), this corresponds to $B_k=0$.   

The RLNs in the pair potential occur at sharp radial positions at $k=k_\Delta$ where $\Delta_k^\lambda=A_k^\lambda=0$. There, the singlet and the triplet  acquire equal strengths, i.e. $\vert\psi_k\vert=\vert F_k\vert$, and their positions depend on the specific non-uniform $k$-dependence of these OPs. In order to capture them in a theoretical model, a full handling of the momentum dependence is necessary\cite{TH_MG}. Reversely thinking, investigating the line nodes in a large range of  anisotropy can provide us with hints in understanding the mechanisms behind the unconventional pairing. We will however focus here on the topological and the thermodynamic properties of NCSs that is valid within a narrow domain of the weakly anisotropic regime. In this work, the weakly anisotropic regime will be characterized by the absence of ALNs.  

In Section. 2 we generalize our previous IS broken self consistent scheme of singlet-triplet mixed-parity pair potentials caused by an arbitrary pairing interaction with the $C_{\infty v}$ symmetry in two dimensions. Section.3 is devoted to the properties of the RLNs. Their topological properties are investigated in Section 3.1 and shown that the relevant topological class is $Z_2$. Section 3.2 is devoted to the low temperature thermodynamic analysis using mainly the energy density of states (DOS) and the specific heat. The basic motivation here is derived from some recent experiments that in some strongly IS broken fully gapped NCSs the thermodynamic data shows BCS-like exponential suppression in low temperatures seemingly pointing at the s-wave pairing. This hints to the fact that, the analysis of such systems can be highly confusing using the thermodynamic data and we believe that this section provides an explanation to this contraversy. The Section 3.3 is devoted to the scattering properties of RLNs at the N-NCS junctions. In this context, we examine the Andreev reflection spectroscopy (ARS) and show that ARS provides a suitable method to capture the distinct signatures of the weakly anisotropic systems.         
     
\section{2. The Model} 
We start with a two dimensional NCS respecting TRS and a general pairing interaction generates the singlet and the triplet components of the pair potential under a SOC. A crucial aspect is that, it is a continuum model which is maximally isotropic and no lattice point group symmetry is assumed. The Hamiltonian in the electronic Nambu-spinor basis $\Psi_{\bf k}^\dagger=(\hat{e}_{{\bf k} \uparrow}^\dagger~\hat{e}_{{\bf k} \downarrow}^\dagger~\hat{e}_{-{\bf k} \uparrow}~\hat{e}_{-{\bf k} \downarrow})$ is given by 
\begin{eqnarray}
{\cal H}=\sum_{\bf k} \Psi_{\bf k}^\dagger {\cal H}_{\bf k} \Psi_{\bf k}= {\cal H}_0+{\cal H}_{soc}+{\cal H}_\Delta 
\label{ham_1}
\end{eqnarray}
where
\begin{eqnarray} 
{\cal H}_{\bf k}=\pmatrix{{\cal H}^0_{\bf k} & \Delta_{\bf k} \cr \Delta_{\bf k}^\dagger & -({\cal H}^0_{-\bf k})^T}\,.
\label{H_matrix} 
\end{eqnarray}
is the $4\times 4$ mean field Hamiltonian with  
\begin{eqnarray}
{\cal H}^0_{\bf k}=\xi_{\bf k}\sigma_0-{\bf G}_{\bf k}.\sigma
\label{H_0}
\end{eqnarray}
describing the kinetic and the SOC parts respectively and 
\begin{eqnarray}
\Delta({\bf k})=\pmatrix{\Delta_{\uparrow \uparrow}({\bf k}) & \Delta_{\uparrow \downarrow}({\bf k}) \cr \Delta_{\downarrow \uparrow}({\bf k}) & \Delta_{\downarrow \downarrow}({\bf k})\cr}=i(\psi_{\bf k}+{\bf d}_{\bf k}.{\bf \sigma})\sigma_y
\label{Delta}
\end{eqnarray} 
is the pair potential. Here, $\xi_{\bf k}=\epsilon_k+\Sigma_{d}({\bf k})$ where $\epsilon_k=\hbar^2 k^2/(2m)-\mu$, $m$ is the band mass, $\mu$ is the chemical potential and ${\Sigma_{d}}({\bf k})$ is the diagonal spin component of the self-energy. Due to the SOC, the off-diagonal contributions can generally arise in the self energy which can be effectively added in the SOC term as ${\bf G}_{\bf k} \to {\bf G}_{\bf k}+\Sigma_{od}({\bf k})$. In the Hartree-Fock mean field approach here, the self energy contributions are ignored. The elements of the OP matrix in Eq.(\ref{Delta}) are given by
\begin{eqnarray}
 \Delta_{\nu \nu^\prime}({\bf k})=-\frac{1}{A}\,\sum_{\bf q}{\cal V}({\bf q})\,\langle \hat{e}^\dagger_{{\bf k+q},\nu}\,\hat{e}^\dagger_{{\bf -k-q},\nu^\prime}\rangle 
 \label{OPs}
 \end{eqnarray}
where ${\cal V}({\bf q})$ is the pairing interaction and $A$ is the sample area. Here, $\Delta_{\uparrow \uparrow}({\bf k})=-\Delta^*_{\downarrow \downarrow}(-{\bf k})$ by the TRS and $\Delta_{\uparrow \uparrow}({\bf k})=F_{\bf k} e^{-i (\phi+\pi/2)}$ by the unitarity of the diagonalization. Furthermore, $F_{\bf k}$ is real and even as discussed in Section.1 and, $e^{-i (\phi+\pi/2)}$ is the phase of the SOC. The excitation spectrum of the Hamiltonian in Eq.(\ref{ham_1}) is given by  
\begin{eqnarray}
E_{\bf k}^\lambda&=&\Bigl[\xi^2_{\bf k}+\vert G_{\bf k}\vert^2+\psi^2_{\bf k}+F^2_{\bf k}+d^2_{z,{\bf k}} \nonumber \\ 
\label{spectrum} \\
&+&2 \lambda \sqrt{(\xi_{\bf k} \vert G_{\bf k}\vert -\psi_{\bf k} F_{\bf k})^2+d^2_{z,{\bf k}}(\vert G_{\bf k}\vert^2+\psi^2_{\bf k})}\Bigr]^{1/2} \nonumber 
\end{eqnarray}
The solution of the general NCS model described by Eq.'s (\ref{ham_1}-\ref{OPs}) requires the fully self consistent calculation of the four order parameters $(\psi_{\bf k}, {\bf d}_{\bf k})$ under a general pairing interaction. 

At this point we emphasize that, one of our motivations in studying the weak anisotropy conditions has something to do with the relation between the $d_{\bf k}$ and ${\bf G}_{\bf k}$. It was shown a long time ago in Ref.\cite{dz} that $d_{\bf k} \parallel {\bf G}_{\bf k}$ yields the thermodynamically most stable configuration with the highest possible $T_c$. It is now a common practice to employ this result in many works. It can be easily seen that, the result in Ref.\cite{dz} becomes exact in the isotropic limit studied here, and satisfied independently from the thermodynamics and the coupling strengths. If the pairing interaction ${\cal V}({\bf q})$ is spin independent, then ${\cal V}({\bf q})={\cal V}(q)$ where $q=\vert {\bf q}\vert$. The physical observables (and particularly the energy spectrum) become independent of the SOC phase $\phi$ which can be defined as a $U(1)$ gauge invariance in the particle-hole sector. An immediate consequence of this is that $d_{z,k}=0$ and $d_{\bf k} \parallel {\bf G}_{\bf k}$.\cite{no_d_z} The Eq.(\ref{spectrum}) can then be expressed in a generalized BCS form as\cite{BCS_form} 
\begin{eqnarray}
E_k^\lambda=\sqrt{(\tilde{\xi}_{k}^{\lambda})^2+(\tilde{\Delta}_k^{\lambda})^2]}
\label{generalized_BCS}
\end{eqnarray}
where $\lambda=\pm$ is the branch index of the broken IS and 
\begin{eqnarray}
\tilde{\xi}_{k}^{\lambda}=\epsilon_{k} +\lambda \gamma_k \vert G_{\bf k}\vert\,\quad {\rm  and} \quad \tilde{\Delta}_k^{\lambda}=(\psi_k -\lambda \gamma_k F_k)
\label{xi_delta}
\end{eqnarray}
are the single particle energy and the momentum dependent pair potential respectively with 
\begin{eqnarray}
\gamma_k=sign(|G_{\bf k}|\epsilon_k-F_k\psi_k)\,.
\label{gamma_k}
\end{eqnarray} 
The energy branches with $\lambda=\pm$ in Eq.(\ref{generalized_BCS}) can have different Fermi surfaces with a different gap opening at the FS as $2\vert \tilde{\Delta}_{k}^{\lambda}\vert$. The bands are in mutual thermodynamical equilibrium by the presence of a single chemical potential, hence the Fermi level can occur at multiple positions in the $k$-space. Together with the nodes of $\tilde{\Delta}_{k}^{\lambda}$, this can give rise to a topological variety. 

The mean field Hartree-Fock solutions of the mixed state OPs in Eq.(\ref{OPs}) can be given in the symmetric form as, 
\begin{eqnarray}
\psi_k=-\frac{1}{A}\,\sum_{k^\prime,\lambda}{\cal V}_s(k, k^\prime)\, \frac{\tilde{\Delta}_{k^{\prime}}^{\lambda}}{4E_{k^{\prime}}^{\lambda}} \Bigl\{f(E_{k^{\prime}}^{\lambda})- f(-E_{k^{\prime}}^{\lambda})\Bigr\}~~~~
\label{Singlet}
\end{eqnarray}
\begin{eqnarray}
F_k=\frac{1}{A}\,\sum_{ k^\prime,\lambda}{\cal V}_t(k, k^\prime)\, \frac{\lambda \tilde{\Delta}_{k^{\prime}}^{\lambda}}{4E_{k^{\prime}}^{\lambda}} \Bigl\{f(E_{k^{\prime}}^{\lambda})- f(-E_{k^{\prime}}^{\lambda})\Bigr\}~~~~
\label{ESP}
\end{eqnarray}
where $ f(x)=1/[exp(\beta x)+1] $ is the Fermi-Dirac factor with $\beta=(k_BT)^{-1}$ as the inverse temperature. The singlet and the ESP-triplet OPs in Eq.'s(\ref{Singlet}) and (\ref{ESP}) are determined by the corresponding interaction channels ${\cal V}_s(k, k^\prime)$ and ${\cal V}_t(k, k^\prime)$. Specifically, $ {\cal V}_s(k, k^\prime)=\langle{\cal V}({\bf |k-k^\prime|})\rangle_a$ and $ {\cal V}_t(k, k^\prime)=\langle{\cal V}({\bf |k-k^\prime|}) \cos{(\phi-\phi^\prime)}\rangle_a$  where $\langle ... \rangle_a$ is the angular average over the relative phase $\phi-\phi^\prime$. In consequence, a bare contact interaction, i.e. ${\cal V}(\vert{\bf k}-{\bf k}^{\prime}\vert)=U$ is insufficient to create pairing in the triplet channel even in the presence of a strong SOC. The term $ {\lambda \tilde{\Delta}_{k^{\prime}}^{\lambda}}/(4E_{k^{\prime}}^{\lambda}) $ in Eq.(\ref{ESP}) is proportional to the difference between the two energy branches. However, a similar term in Eq.(\ref{Singlet}) represents the sum of the same contributions in $\psi_k$. A non-local pairing interaction and the SOC are therefore essential factors in the $k$-dependence of the OPs in the mixed state. This affects most importantly the RLN positions, the topology of the energy bands and the low temperature properties as discussed below. 

\section{3. The Radial Line Nodes}
In general, whether point or line, the nodes can be present in two distinct levels: a) the pair potential $\tilde{\Delta}_k^{\lambda}$ and b) the energy spectrum $E_k^\lambda$. 

In strongly anisotropic NCSs the singlet and the triplet configurations are permitted by the crystal point group symmetries. This allowed the development of group theoretical models in the general classification of the ALNs and the energy band topology. A recent review\cite{general_nodes} can be very useful for a complete summary in this respect. On the other hand, in the weakly anisotropic NCSs the point groups are ineffective and the ALNs are absent. However, these materials can still display puzzling low temperature behaviour\cite{NCS_review2,swave_NCS1}, i.e. some NCSs with strong spin-orbit coupling display a fully gapped s-wave behaviour in thermodynamical response. In this work we bring an explanation to this conflict and show that the RLNs in the pair potential can simulate an isotropic s-wave superconductor. A fully gapped spectrum is present when the RLN positions are away from the Fermi surface, that is the case (a) above. More precisely, these type of nodes are topologically classified according to their position with respect to the Fermi surface. They can be most accurately identified by the ARPES\cite{ARPES}, Andreev Reflection Spectroscopy\cite{Andreev_conductance} or other ingenious measurements\cite{Kurter}. 

If an RLN is on the Fermi surface, there is a gappless spectrum and the case (b) occurs, i.e. the energy nodes in $E_k^\lambda$. Because of this additional Fermi surface matching condition, these nodes are physically rare, but when they occur, they dominate the low temperature thermodynamic behaviour\cite{weak_anisotropy}. Hence, a concise analysis of RLNs is necessary, which we do next.  

\subsection{\it 3.1 The Topology of the RLNs}
In this section we demonstrate that the RLNs' topology is encoded in the position of the nodes and their nodal positions can, in principal, be externally controlled. This result is important from the future device applications, as experimental progress in this direction can lead us to the topologically controllable systems. The angular line nodes are, on the other hand, much less flexible externally due to the dominant effect of the crystal symmetries. We now carry on a topological analysis using two different methods and demonstrate that the topology is determined by the positions of the RLNs. 

a) {\it The block-diagonal Hamiltonian method}: In the absence of the $d_{z,{\bf k}}$ type pairing, the mixed state Hamiltonian in Eq.(\ref{H_matrix}) can be block-diagonalized in the SOC eigenbasis $ (\hat{a}_{{\bf k} +}^\dagger~\hat{a}_{{\bf k} -}^\dagger~\hat{a}_{-{\bf k} +}~\hat{a}_{-{\bf k} -})$ where $ \hat{a}_{{\bf k} \Lambda}=\frac{1}{\sqrt{2}}(\hat{e}_{{\bf k} \uparrow}+ \lambda \gamma_k e^{i \phi} \hat{e}_{{\bf k} \downarrow}) $ with $ \lambda=\pm $. Each block is described by a $2\times 2$ matrix in the form written by $H^{\lambda}={\bf h}_{\bf k}^{\lambda}.{\bf \Lambda}$ with ${\bf \Lambda}$ as the Pauli  matrices in the SOC basis and ${\bf h}_{\bf k}^{\lambda}=(h_x^\lambda,h_y^\lambda,h_z^\lambda)=(\tilde{\Delta}_k^{\lambda} \cos\phi,-\tilde{\Delta}_k^{\lambda} \sin\phi,\tilde{\xi}_{k}^{\lambda})$. One way to examine the energy band topology of RLNs is then to investigate each block-diagonal branch separately by the two-dimensional mapping $(k_x,k_y) \to {\hat n}_{\bf k}^{\lambda}$ where ${\hat n}_{\bf k}^{\lambda}={\bf h}_{\bf k}^{\lambda}/E_{\bf k}^{\lambda}$ is the Hamiltonian unit sphere. This two-dimensional mapping is described by the Chern number
\begin{eqnarray}
N_1^{\lambda}=\frac{1}{8\pi}\int\,d^2k\,\epsilon_{ij}\,{\hat n}_{\bf k}^{\lambda}.\Bigl(\frac{\partial{\hat n}_{\bf k}^{\lambda}}{\partial k_i} \times \frac{\partial{\hat n}_{\bf k}^{\lambda}}{\partial k_j} \Bigr)
\label{chern_index}
\end{eqnarray} 
where $\epsilon_{ij}$ with $i,j=x,y$ is the antisymmetric tensor. The two branches can have independent indices given by the winding of ${\bf h}_{\bf k}^{\pm}$ on $S_2$. 

\begin{figure}
\includegraphics [scale=0.52]{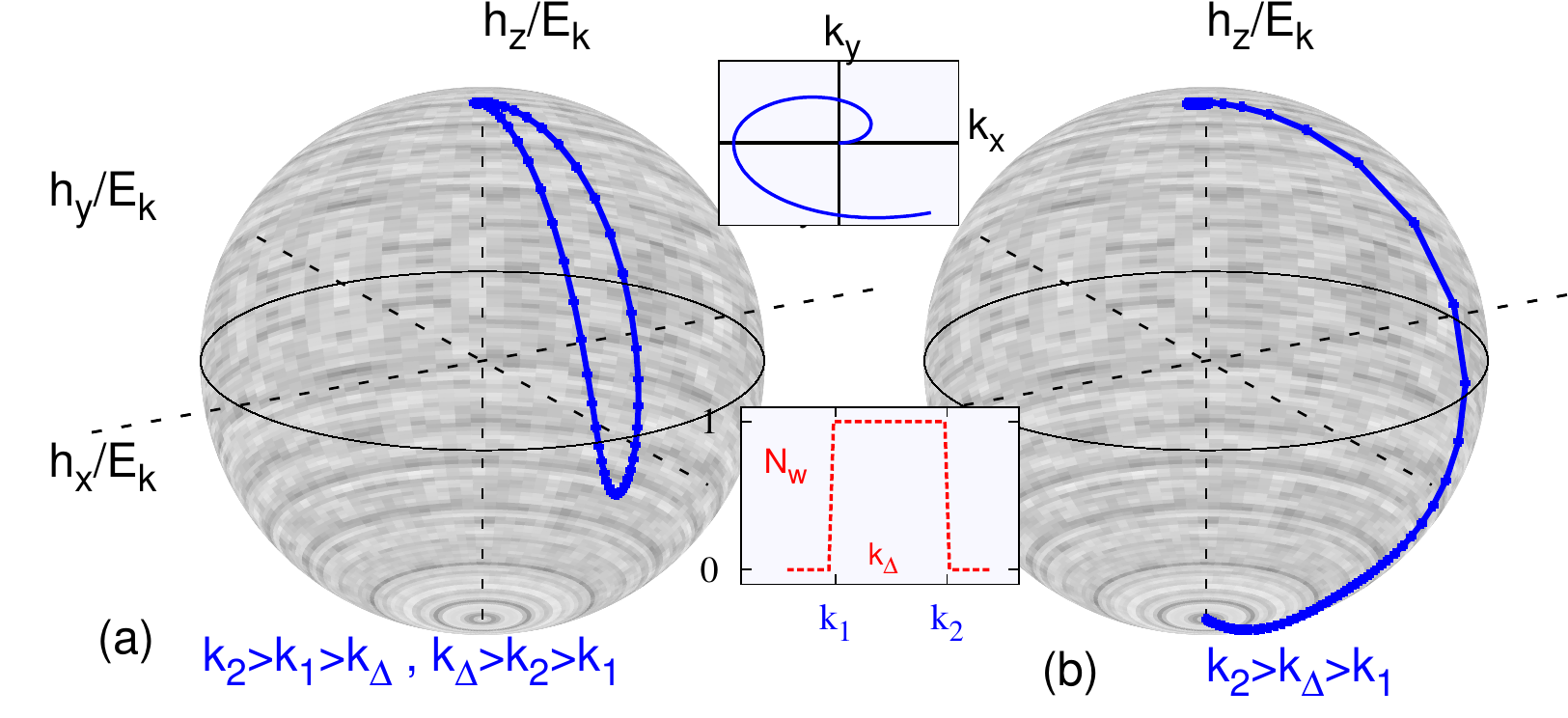}\\
\vskip0.2truecm
\includegraphics [scale=0.33]{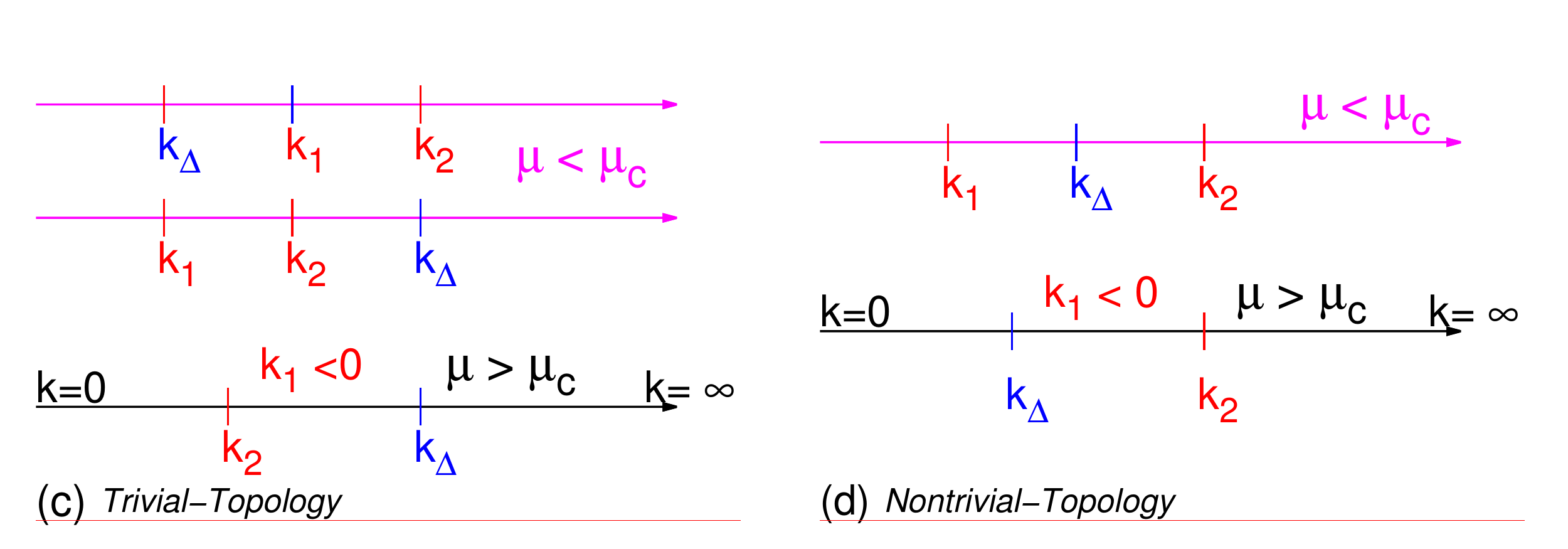}
\vskip-0.2truecm
\caption{(Color online) Nodal positions of $\tilde{\xi}_k^{\lambda}$ and $\Delta_k^{\lambda}$ when $\lambda=+$ depicted respectively as $k_1^+, k_2^+$ and $k_\Delta^+$ with different topologies as indicated in a) as trivial $ N_1^+=0 $ , b) as nontrivial, $ N_1^+=1$. The zeros $k_{1,2}^+$ are determined by $\mu$ and $\alpha$. The topology is illustrated on the unit sphere as a) trivial for case (c), and b) nontrivial for case (d).}  
\label{unit_sphere}
\end{figure}

The complete topological classification is made once all distinct configurations of the nodes in $\tilde{\Delta}_k^{\pm}$ relative to the position(s) of the Fermi level are identified. For this, we start with the {\it kinetic} term in the BCS-like form in Eq.(\ref{generalized_BCS}) given by
\begin{eqnarray}
\tilde{\xi}_k^{\lambda}=\hbar^2 (\gamma_k k- k_1^\lambda)(\gamma_k k- k_2^\lambda)/(2m)
\label{kinetic1} 
\end{eqnarray} 
Here $k_1^\lambda, k_2^\lambda$ are the zeros of $\tilde{\xi}_k^{\lambda}$. A positive $k_j^\lambda$ is a Fermi momentum on $j$'th Fermi surface of the corresponding branch. We assume that $ k_2^\lambda >  k_1^\lambda$. For the moment, we take $\gamma_k=1$ and discuss its effect later. The Fermi wavevectors for the $+$ branch are,  
\begin{eqnarray}
k_2^+=&\frac{m}{\hbar^2}\Bigl[-\alpha+\sqrt{\alpha^2+2\frac{\hbar^2}{m}\mu}\Bigr] \label{energy_nodes_1} \, \qquad &(\mu > 0) \\ 
k_{1,2}^+=&\frac{m}{\hbar^2}\Bigl[\alpha \mp \sqrt{\alpha^2+2\frac{\hbar^2}{m}\mu}\Bigr] \label{energy_nodes_2}  \, \qquad &(\mu < 0)
\end{eqnarray}
Here $\mu=-\hbar^2 k_1^\lambda k_2^\lambda/(2m)$ and $\alpha=-\lambda \hbar^2(k_1^\lambda+k_2^\lambda)/(2m)$ are the physical parameters which can be used to vary the $k_{1,2}^\lambda$. All five possibilities are shown in Fig.\ref{unit_sphere} for the $+$ branch. The $-$ branch is analyzed similarly. 

Concentrating on the $+$ branch, we will assume that the triplet-to-singlet ratio $\vert F_k/\psi_k \vert$ can have values smaller and larger than unity in different $k$ regions. The pair potential is then allowed to have a node, let's say at $k_\Delta^+$ of the $+$ branch, and $\Delta_k^{+}=\delta_+(k)(k-k_\Delta^+)$ where $\delta_+(k)$ is a smooth function representing the other (irrelevant) details. At $k=k_\Delta^+$ we have that $\vert \psi_{k_\Delta^+}\vert =\vert F_{k_\Delta^+}\vert$. It will also be assumed that there is only a single position where such a $k_\Delta^+$ exists. In general there is nothing to prevent the number of such points to be larger than one and a complete map of such details is an additional asset to the understanding of the pairing mechanism. Particularly, a mathematical inversion of Eq.'s(\ref{Singlet}) and (\ref{ESP}) can be useful to get information about the pairing interactions by knowing the nodes of $\Delta_k^{\lambda}$. We nevertheless avoid this interesting idea to a later work and confine our attention to maximally one RLN at $k_\Delta^\lambda$ for each $\lambda$.   

In the pure triplet state\cite{pure_triplet_class}, $\psi_{\bf k}=0$ (in general realized either the TRS invariant helical p-wave or the TRS breaking chiral p-wave). In this case, the topology of the superconducting bands is decided by $\mu$ only. The Chern index in Eq.(\ref{chern_index}) is an integer yielding 
\begin{eqnarray} 
N_1^{+}(\mu)=\cases{0 & $\quad$ {\rm for} $\quad \mu<0$ \cr 1 & $\quad$ {\rm for} $\mu>0$} 
\label{chern_index2}
\end{eqnarray}
This picture is quite similar to the $Z_2$ topology of the two dimensional QSHI, one dimensional polyacetylene\cite{SSH} and the spinless one dimensional p-wave superconductor\cite{Kitaev}. 

In the mixed state however, singly-parameterized characterization is not sufficient. Moreover, Eq.(\ref{chern_index}) is not integer-valued. The first case is remedied by a doubly-parameterized characterization. Due to the additional spin-orbit degree of freedom, the Chern index in Eq.(\ref{chern_index}) depends on $\mu$ and $\alpha$, i.e. $N_1^{\lambda}(\mu,\alpha)$. The positions of $k_1^\lambda, k_2^\lambda$ relative to $k_\Delta^\lambda$ can be used to classify the topology using Eq's.(\ref{energy_nodes_1}), (\ref{energy_nodes_2}) and the like for $\lambda=-$. 

We again concentrate on the $+$ branch. If $\mu<0$ the kinetic term can have two Fermi wavevectors $k_1^+, k_2^+ >0$ given by Eq.(\ref{energy_nodes_2}), or none, i.e. $k_1^+, k_2^+ <0$, whereas for $\mu>0$ there is one Fermi momentum, i.e. $k_2^+$ [as given by Eq.(\ref{energy_nodes_1})]. To begin, one can start from a trivial configuration such as $k_1^+<k_2^+<k_\Delta^+$ which is then used as a reference for all other topological configurations. The mapping ${\bf k} \to {\hat n}_{\bf k}^\lambda$ is described in Fig.\ref{unit_sphere}.(a) and (b). Here, $k\to \infty$ corresponds to the north pole $(0,0,1)$ of the Hamiltonian unit sphere. The trivial and nontrivial topologies for five distinct configurations of $k_1^\lambda, k_2^\lambda$ and $k_\Delta^\lambda$ are also indicated in Fig.\ref{unit_sphere}.(c) and (d). These configurations differ in topology by the number of Fermi level crossings of the node $k_{\Delta}^\lambda$ where the topology is changed by every single crossing. For $k_1^+<k_\Delta^+<k_2^+$, $k_\Delta^+$ is mapped to the south pole $(0,0,-1)$, whereas for $k_\Delta^+<k_1^+<k_2^+$, it is mapped to the north pole $(0,0,1)$. 

The non-integer valued index can be remedied by considering the reduced integral range $k_{\Delta}^\lambda\le k < \infty$. The results thus obtained from Eq.(\ref{chern_index}) are shown in the inset of Fig.\ref{unit_sphere}. 

This concludes the investigation of the block diagonal formalism. We now present another method by bringing the same Hamiltonian into a block non-diagonal form.     

b) {\it The block-nondiagonal Hamiltonian method}: A method for the topological index was proposed in Ref.'s\cite{NCS_nodal_topo0,NCS_nodal_topo,NCS_nodal_topo2} for Hamiltonians respecting "chiral symmetry" which is given by the product of the TRS and the particle hole symmetry. Both symmetries are preserved in our case here. In systems with chiral sysmmetry a new way of defining topological index can be developed by bringing the Hamiltonian in Eq.(\ref{ham_1}) into the block off-diagonal form. In our case, this is obtained by a global unitary transformation $V$ acting on ${\cal H}_{\bf k}$ in Eq.(\ref{H_matrix}) as \cite{NCS_nodal_topo0,NCS_nodal_topo,NCS_nodal_topo2}  
\begin{eqnarray}
V{\cal H}_{\bf k}V^\dagger=\pmatrix{0 & D_{\bf k} \cr 
  D_{\bf k}^\dagger & 0} ,\quad   V=\frac{1}{\sqrt{2}}\pmatrix{\sigma_0 & -\sigma_2 \cr 
  i \sigma_2 & i \sigma_0} ~~~~
\label{off-diag1}
\end{eqnarray} 
where $D_{\bf k}=C_k[cos(\phi_k) \sigma_z+ i sin(\phi_k)\sigma_0]-i B_k \sigma_2$  with $C_k=|{\bf G_k}|-i F_k$ and $B_k=\xi_k+i \psi_k$. Similarly to Eq.(\ref{chern_index}), here $D_{\bf k}$ is well defined only in those ${\bf k}$ points where the energy spectrum Eq.(\ref{generalized_BCS}) is nonvanishing, i.e. when the gap is full. In Ref.'s\cite{NCS_nodal_topo0,NCS_nodal_topo,NCS_nodal_topo2,NCS_nodal_topo3} a {\it momentum-dependent} (contrary to global) topological index is defined as
\begin{eqnarray}
N_{2}(k_\perp)=\frac{1}{2\pi} \Im m\Bigl\{ \int_{-\infty}^\infty dk_\parallel \, \partial_{k_\parallel} \ln det({\tilde{D}_{\bf k}}) \Bigl\}
\label{off-diag3}
\end{eqnarray} 
where $k_\parallel$ and $k_\perp$ are coordinates fully parametrizing the ${\bf k}$-plane. Here we transformed $ D_{\bf k}\rightarrow {\tilde{D}_{\bf k}}$ as $det({\tilde{D}_{\bf k}})=det(D_{\bf k})/|det(D_{\bf k})|$. 
 
In the context of this work, $N_{2}$ is, desirably, a global index, due to the $\phi$-independence of $det({\tilde{D}_{\bf k}})$. For the same reason, $k_\parallel$ integral can be split into a pair of equivalent radial integrals, i.e. $k_\parallel=k$ at $\phi=0$ and $\phi=\pi$. The Eq.(\ref{off-diag3}) can then be turned into  
\begin{eqnarray}
N_{2}=\frac{1}{\pi} \Im m\Bigl\{ \int_{0}^\infty dk \, \partial_{k} \ln det({\tilde{D}_{\bf k}}) \Bigl\}
\label{off-diag2}  
\end{eqnarray}
The Eq.(\ref{off-diag2}) can now be shown to be connected with $N_1^\lambda$ in Eq.(\ref{chern_index}), i.e. $N_2=N_1^++N_1^-$. 

For this, we use $D_{\bf k}, C_k$ and $B_k$ as defined below Eq.(\ref{off-diag1}) to find that $det({D_{\bf k})}=({\tilde \xi}_k^+ +i{\tilde \Delta}^+_k)({\tilde \xi}_k^- +i{\tilde \Delta}^-_k)$. Therefore the Eq.(\ref{off-diag2}) is
\begin{eqnarray}
N_{2}&=&\frac{1}{\pi} \sum_{\lambda} \int_0^\infty dk \,\partial_k [arg({\tilde \xi}_k^\lambda +i{\tilde \Delta}^\lambda_k)]~.
\label{winding_mixed2}
\end{eqnarray}
Since $arg({\tilde \xi}_k^\lambda +i{\tilde \Delta}^\lambda_k)$ is the polar angle $\theta^\lambda=tan^{-1}{\tilde \Delta}^\lambda_k/{\tilde \xi}_k^\lambda$ of the Hamiltonian unit vector $n^\lambda(\theta,\phi)=({\tilde \Delta}_k^\lambda \cos\phi,{\tilde \Delta}_k^\lambda\sin\phi,{\tilde \xi}_k^\lambda)$, the Eq.(\ref{winding_mixed2}) is identical with the winding of the polar angle on the unit circle at a fixed longitude $\phi^*$, i.e. $N_2=\sum_{\lambda} \int d\theta^\lambda/\pi$. There is therefore a one-to-one correspondence between Eq.(\ref{chern_index}) and Eq.(\ref{winding_mixed2}) [hence Eq.(\ref{off-diag2})].  

In order to obtain an integer index from Eq.(\ref{off-diag3}), the consideration of the reduced range is the simplest. Another alternative technique was also suggested. Assume that the pair potential is sufficiently weak near the Fermi surface. One can use the positions of the multiple sectors of the Fermi surface and linearly expand $\tilde{\xi}_k^\lambda$ and $\tilde{\Delta}_k^\lambda$ around the $i$'th Fermi surface. The expectation is that a continuous deformation assumed in the linear expansion does not change the topology, hence a discrete index is expected. It was shown in Ref.'s\cite{NCS_nodal_topo0,NCS_nodal_topo,NCS_nodal_topo2,NCS_nodal_topo3} that Eq.(\ref{off-diag3}) can then be written as    
\begin{eqnarray}
N_{2}=-\frac{1}{2}\sum_{k_i} && sign[\partial_k ({\tilde{\xi}_k^+} {\tilde{\xi}_k^-})|_{k=k_i}] \nonumber \\ 
&& \times sign[({\tilde{\Delta}_k^+ }{\tilde{\xi}_k^-}+{\tilde{\Delta}_k^- }{\tilde{\xi}_k^+} )|_{k=k_i}]
\label{winding_mixed}
\end{eqnarray} 
where point(s) $ k_i^\lambda$ are the Fermi momenta given by $ \tilde{\xi}^\lambda_k|_{k_i^\lambda}=0 $. This can be written as a sum of separate branches as, 
\begin{eqnarray}
N_{2}=-\frac{1}{2}\sum_{\lambda}\,\sum_{\tilde{\xi}^\lambda_{k_i}=0} sign[\partial_k {\tilde{\xi}_k^\lambda} |_{k=k_i}] sign[{\tilde{\Delta}_k^\lambda }|_{k=k_i}]
\label{winding_mixed1}
\end{eqnarray}  
The $N_2^\lambda$ calculated from Eq.(\ref{winding_mixed1}) yields the same result as $N_1^\lambda$ using the reduced integration range [see Fig.(\ref{unit_sphere})].  
With the equivalence of both methods in parts (a) and (b), we revisit Fig.\ref{unit_sphere} (c) and (d) for a summary. The topological indices defined in this section are undefined when the spectral gap closes. The topology of the RLNs in the full gap configuration requires that $k_\Delta^\lambda$ is not on the Fermi surface, i.e. $k_\Delta^\lambda \ne (k_1^\lambda,k_2^\lambda)$. The topology can then be classified by the $Z_2$ index according to the position of the RLN in the pair potential with respect to the Fermi surface. This requires that the configurations where $k_\Delta^\lambda=k_1^\lambda$ or $k_\Delta^\lambda=k_2^\lambda$ are topologically undefined. Since these specific configurations are where the energy gap closes, we have the important result that, an RLN in the energy spectrum occurs at the boundary of two distinct topological regions. This concludes our discussion on the topology of the RLNs. We now investigate the influence of the RLNs on the thermodynamic observables where the results of the Section 3.1 will also be used. 

\subsection{\it 3.2 Thermodynamic Signatures}
In fully gapped NCSs, specific heat, penetration depth  and other thermodynamic observables display exponential suppression in temperature in sufficiently low temperatures. For this reason, it is difficult in thermodynamic experiments to separate the unconventional pairing in these systems from the fully gapped trivial s-wave superconductors.

On the other hand, the gapless superconductors -mostly studied in the context of ALNs in the anisotropic regime- can be easily identified in thermodynamical measurements with their distinct scaling behaviour near vanishing excitation energies. In this case, the exponential suppression in temperature is replaced by a clean power law depending on the nodal dimensions. It is known that in two dimensional systems, the point nodes can yield in the specific heat a $T^3$, whereas the ALNs yield a $T^2$ dependence\cite{NCS_review}. 
\begin{figure*}[ht] 
\includegraphics [width=80mm,height=45mm]{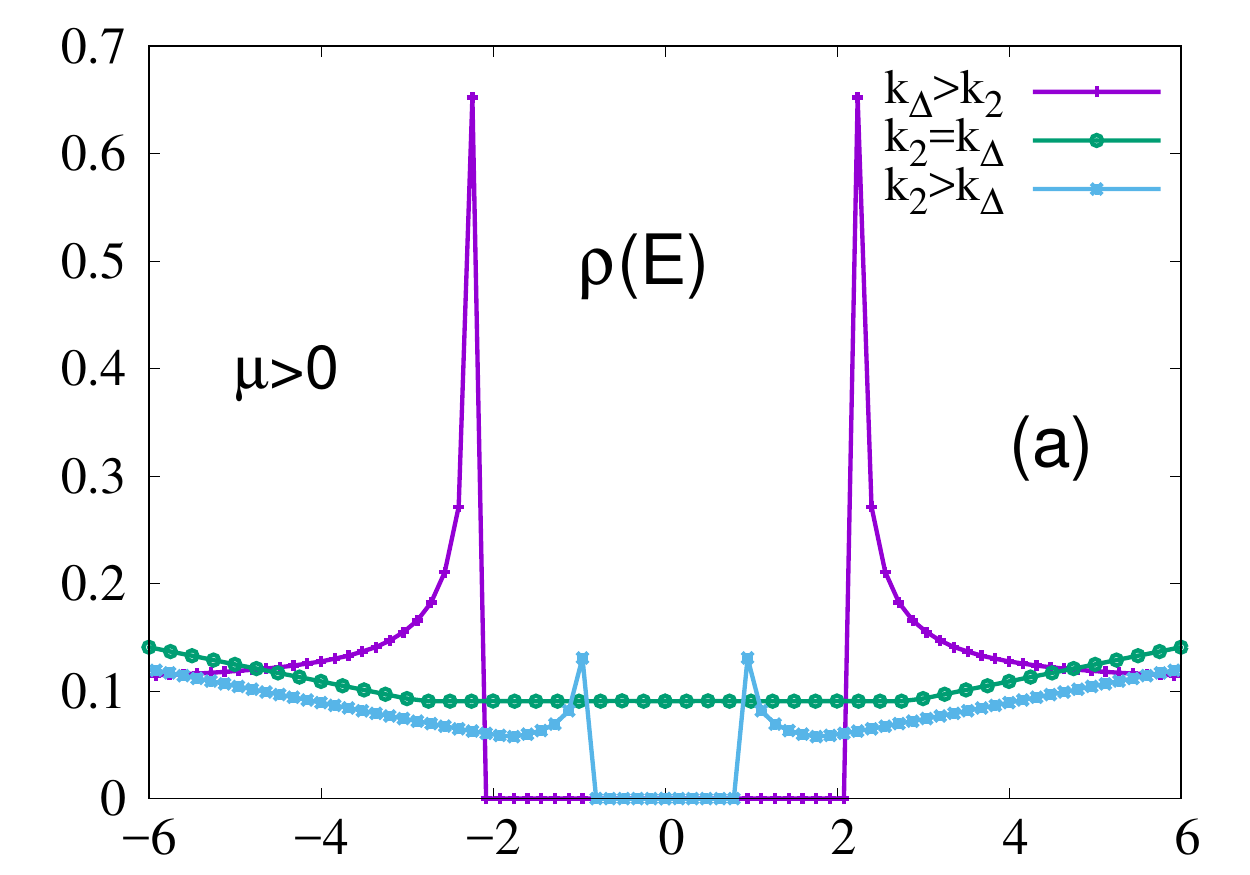}
\includegraphics [width=80mm,height=45mm]{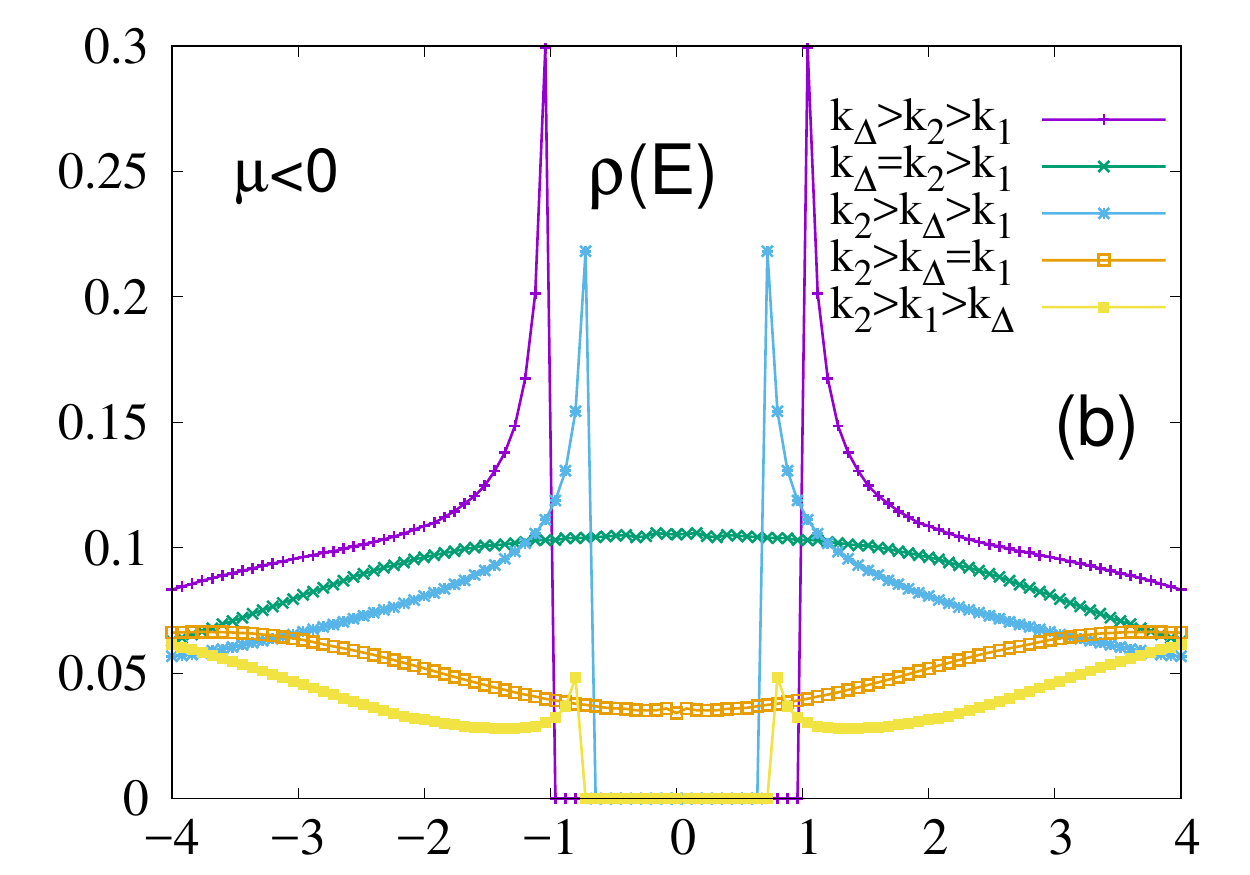} \\
\vskip0.2truecm
\includegraphics [width=80mm,height=45mm]{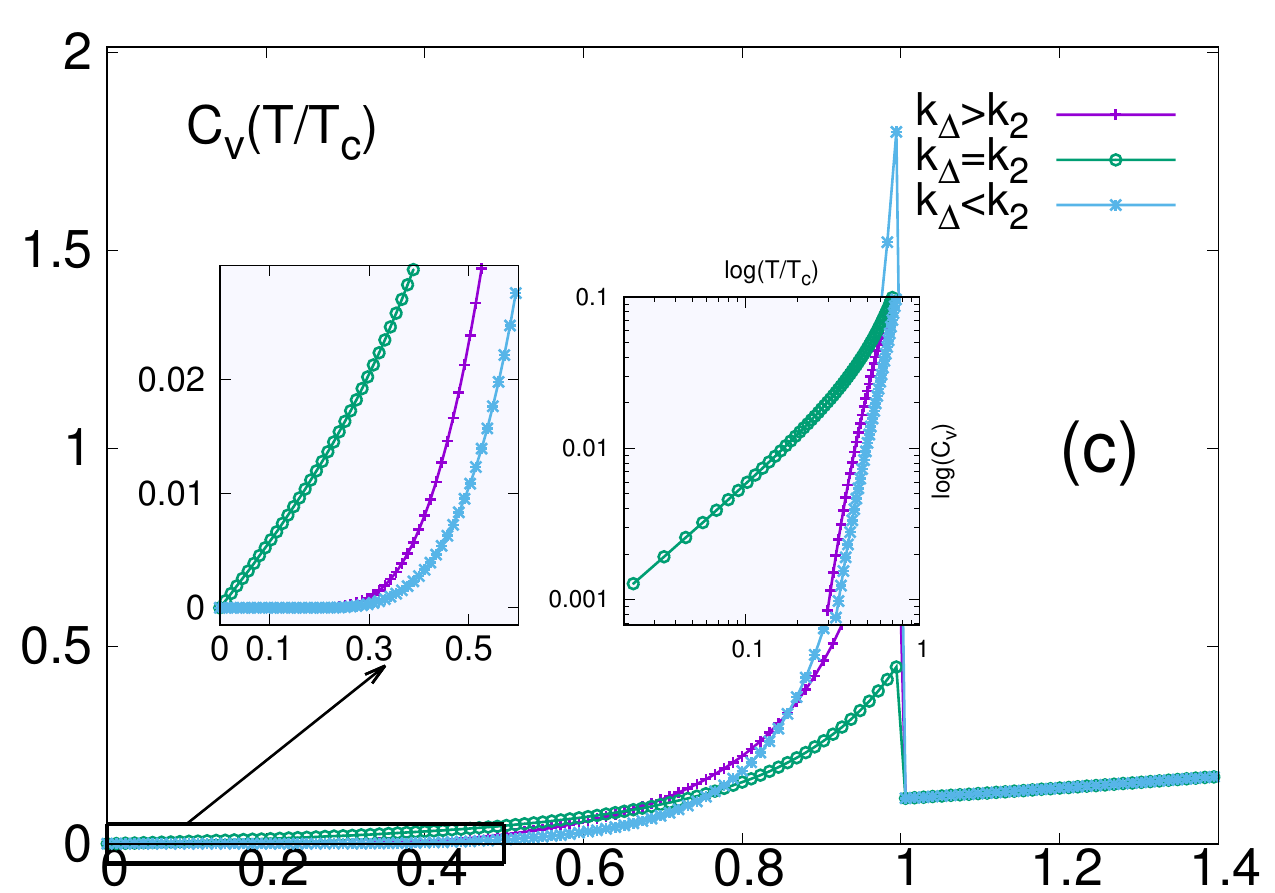}
\includegraphics [width=80mm,height=45mm]{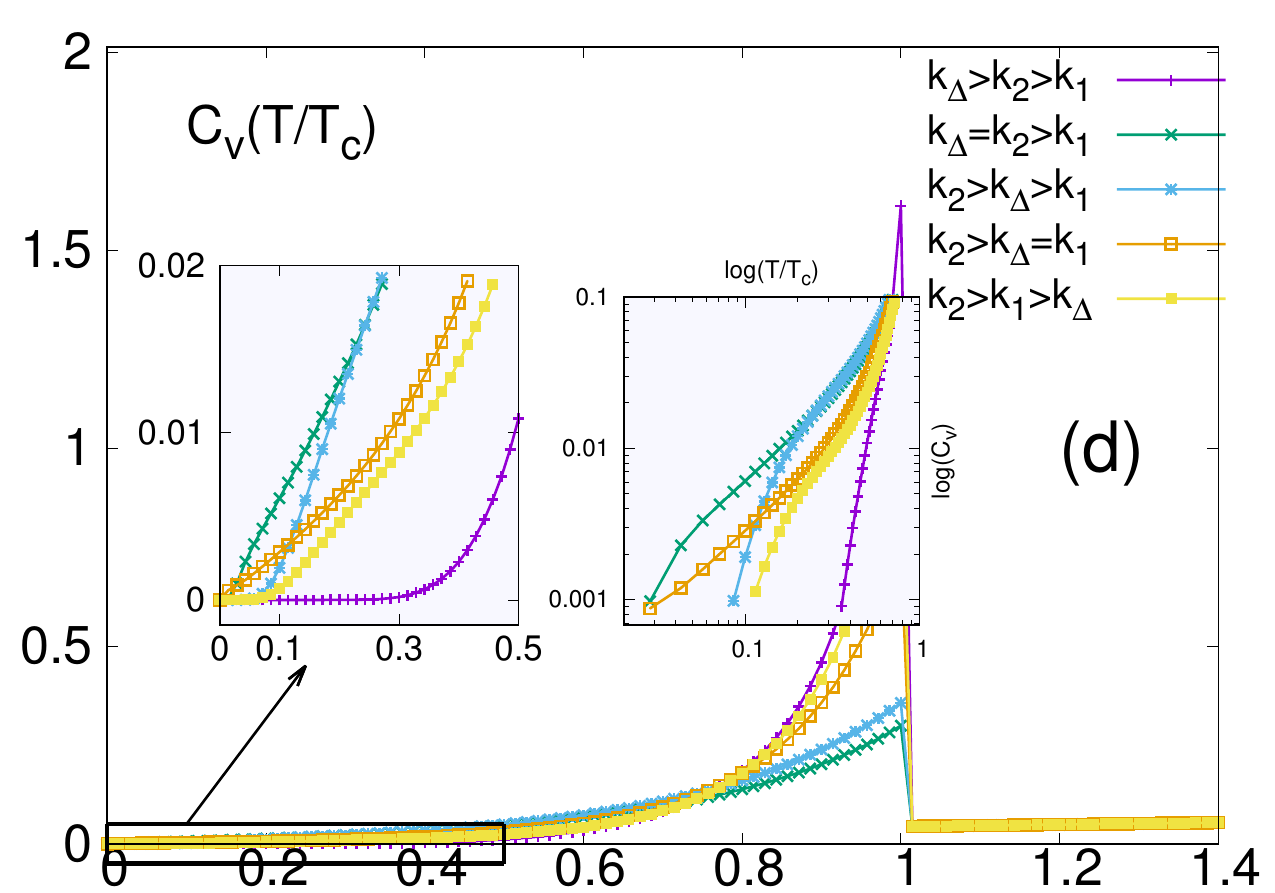}
  \label{FP}\caption{(Color online) The effect of the Fermi level crossing of the node $k_\Delta$ in the pairing potential for $\mu>0$ in a) the DOS $\rho(E)$ and b) the $C_V$ corresponding to the cases $k_\Delta < k_2$, $k_\Delta = k_2$ and $k_\Delta > k_2$. The effect of the Fermi level crossing of the energy gap node $k_\Delta$ for $\mu<0$ on the a) $\rho(E)$ and b) $C_V$ corresponding to 5 different positions of $k_\Delta$ color coded in (b), as also indicated in Fig.\ref{unit_sphere}.(a,b). The insets magnify the low $E$ and low $T$ region of $\rho(E)$ and $C_V$ which are linear for $k_\Delta=k_1$ and $k_\Delta=k_2$.Color coding applies to both figures. }
\label{transition_to_from_scaling}
\end{figure*}   
 
It is our goal in this section to show that, the weakly anisotropic systems where ALNs are not present, behave thermodynamically like the ordinary s-wave superconductors. This is so even in the presence of strongly mixed singlet-triplet components with RLNs present in the pair potential. In order to study the thermodynamics of these systems, we start with the energy DOS of the branch $\lambda$, 
\begin{eqnarray}
\rho_\lambda(E)=\int \frac{d{\bf k}}{(2\pi)^2}\,\delta(E-E_{\bf k}^\lambda)
\label{DOS1}
\end{eqnarray}
and examine its behaviour in the context of Sec.3.1. We consider $E_k^\lambda$ in the context of Eq.(\ref{generalized_BCS}) and (\ref{kinetic1}) also allowing the pair potential to have a simple RLN at $k_\Delta^\lambda$, i.e. $\Delta_k^\lambda \simeq b_\lambda(k-k_\Delta^\lambda)$. If we concentrate on the region $k \simeq k_2^\lambda$ for a fixed $\lambda$ and $\mu > 0$, then $\tilde{\xi}_k^\lambda \simeq a_\lambda (k-k_2^\lambda)$. Here $a_\lambda$ and $b_\lambda$ are some coefficients. We find that
\begin{eqnarray}
\rho_\lambda(E)=\frac{1}{2\pi}\,\frac{E}{a_\lambda^2(1-k_2^\lambda/k)+b_\lambda^2(1-k_\Delta^\lambda/k)}\Big\vert_{k=k_\lambda(E)}~~~~
\label{DOS2}
\end{eqnarray}
and $k_\lambda(E)$ is where $E^\lambda_k=E$. The Eq.(\ref{DOS2}) indicates that, for large energies $\rho(E) \sim E$. The small energy limit of DOS depends on whether a zero energy mode at a finite $k$ is supported in the spectrum. For the zero energy mode $k_2^\lambda=k_\Delta^\lambda$ must be physically realized, i.e. the node in the pair potential must occur at the Fermi level. In this case, Eq.(\ref{DOS1}) implies that in the vicinity of the zero mode $\rho_\lambda(E)=k_E^\lambda/(2\pi\sqrt{a_\lambda^2+b_\lambda^2})$, i.e. a constant. On the other hand, if $k_2^\lambda \ne k_\Delta^\lambda$, there is a gap in the spectrum for $E < E_{min}^\lambda=a_\lambda b_\lambda \vert k_2^\lambda-k_\Delta^\lambda\vert/\sqrt{a_\lambda^2+b_\lambda^2}$ with a divergent DOS at the gap edge. The DOS for this $\mu >0$ case is summarized in Fig.\ref{transition_to_from_scaling}.(a). Before commenting on this case, we examine the DOS for $\mu<0$. Here, there are two Fermi surface positions or none. Let us assume that there can be one RLN of the pair potential  at $k_\Delta^\lambda$ for each branch $\lambda$. In this case, there can be two, one or zero number of energy nodes and the picture obtained for the $\mu>0$ case in the DOS is repeated here according to the number of energy nodes. Finally, the DOS for $\mu<0$ is shown in Fig.\ref{transition_to_from_scaling}.(b). 

In Fig.'s \ref{transition_to_from_scaling}.(a) and (b), the behaviour of the DOS in the fully gapped regimes is unseparable from a conventional s-wave superconductor. Different behaviour from the conventional superconductor appears when the node is located on the Fermi surface. This corresponds to an RLN in the energy spectrum and in contrast to the ALNs where $\rho(E) \sim E^\nu$ with $\nu$ being an integer exponent depending on dimensionality\cite{NCS_review}, here the DOS acquires a constant value. A comparison with the previous section shows that, the point where the discontinuous jump occurs in $\rho(0)$ is a boundary between topologically two distinct regimes. The experimental observation of this discontinuity should be considered as a significant evidence about the presence of RLNs and, any thermodynamic quantity based on $\rho(E)$ is expected to have this signature. For instance the specific heat given by  
\begin{eqnarray}
C_V(T)=\sum_\lambda \int dE \rho_\lambda (E) E \frac{f(E)}{dT}
\label{C_v}
\end{eqnarray}
where $f(E)$ was define before, displays a sharp transition from the exponential suppression to the linear dependence as shown in Fig.\ref{transition_to_from_scaling}.(c) and (d). The temperature dependence of the $C_V$ in an NCS with RLNs is therefore very similar to that of the $s$-wave BCS superconductor. This is a crucial information which may be useful in resolving some of the experimental controversies. Indeed, recently a number of thermodynamic experiments were reported on NCSs with strong IS breaking\cite{swave_NCS1} and the list is rapidly extending\cite{swave_NCS}. In these works, the thermodynamic data is similar to Fig.\ref{transition_to_from_scaling}.(c) and (d) and the opinion of those authors is in favour of the conventional s-wave BCS superconductivity. On the other hand, other evidences were also emphasized therein pointing at the unconventional pairing.    

Our results in this section can demonstrate that a fully gapped superconductor with an RLN located off the Fermi surface can display thermodynamic data like ordinary s-wave superconductors at the same time being topologically unconventional in a strongly mixed state.            

An important side remark is that, if $\gamma_k$ changes sign between the two Fermi wavevectors $k_1$ and $k_2$, then both gaps $\Delta_k^{\pm}$ are allowed to have RLNs. This case is interesting but certainly a very rare circumstance. Its experimental identification may be difficult to reveal in thermodynamic measurements, but it may be possible by ARS which we discuss next.   

\subsection{\it 3.3 The Andreev Reflection Spectroscopy with weakly anisotropic NCSs}
The arguments raised above show that the thermodynamic data can be misleading in understanding the OP symmetry. We furthermore demonstrated that, the RLNs offer an explanation to these controversies and they are most likely present in weakly anisotropic systems. 

We can differentiate the strongly anisotropic conditions from the weakly anisotropic ones, for instance as shown in the top and the bottom slices of Fig\ref{ALN_to_RLN}.(b), by using probes that can control the energy and the momentum vector at the single particle level. The Andreev conductance (AC) measurements have been useful experimental tools for obtaining information about the pairing symmetry of the s, d and chiral p-wave superconductors.\cite{NCS_review,TMYYS} In this section, we demonstrate that the AC can be also an efficient probe for the weakly anisotropic ones.  

We consider the junction of a normal metal (N) with a weakly anisotropic NCS in the $x-y$ plane and calculate the AC at an N-NCS interface where $x<0$ is the N side without SOC and $x>0$ is the NCS with SOC. We assume ideal conditions with no interface potential which allows us to discard the normal reflection\cite{BTK}. 

Initially, an electron, spin-polarized in the z-direction, is sent normal to the N-NCS interface from the N side at the wavevector $K_e=\sqrt{2m(E+\mu_N)/\hbar^2}$ where $E$ is the energy of the incident electron and $\mu_N=\hbar^2 k_F^2/(2m)$ is the chemical potential in the N region with $k_F$ being the Fermi wavevector in the N side. We perform our calculations for the zero voltage bias, i.e. $\mu_N=\mu_{NCS}$. The corresponding wavefunctions are \cite{AR_NCS}          
\begin{eqnarray}
\Psi_N(x)&=&\{e^{iK_e x} (1, 0, 0, 0) ^T +a\,e^{iK_h x} (0, 0, 1, 0)^T \nonumber \\ 
&+&b\,e^{i K_h x} (0, 0, 0, 1)^T +c\,e^{-iK_e x} (1, 0, 0, 0)^T \nonumber \\ 
&+& d\,e^{-iK_e x} (0, 1, 0, 0) ^T \} \qquad {\rm and}
\label{wf1} \\
\Psi_{S}(x)&=&\{ c_1 \,e^{iq_1^+x} ( u_1, \eta \,u_1, \eta^* \,v_1, v_1 ) ^T \nonumber \\ 
&+&c_2 \,e^{iq_2^+x} (-\eta^* \,u_2, u_2, v_2, -\eta^* \,v_2) ^T \nonumber \\ 
&+& c_3 \,e^{-iq_1^-x} (-\eta^* v_1, v_1, u_1, -\eta^* \,u_1)^T \nonumber \\
&+& c_4 \,e^{-iq_2^-x} (v_2, \eta \,v_2, \eta \,u_2, u_2)^T\} \nonumber
\end{eqnarray}
where $\eta=e^{i \phi}$ evaluated at the corresponding transmitted wavevector $q_{1 \choose 2}^\pm$ in Eq.'s(\ref{wf1}), and for $\vert E\vert > \vert \Delta_\pm \vert$
\begin{eqnarray}
u_{1 \choose 2}&=&\frac{1}{2} \sqrt{\bigg(1+\sqrt{1-\Big(\frac{\Delta_\pm}{E} \Big)^2}\bigg)} \nonumber\\
\label{u_and_v_1}\\
v_{1 \choose 2}&=&\frac{1}{2} \sqrt{\bigg(1 - \sqrt{1-\Big(\frac{\Delta_\pm}{E} \Big)^2}\bigg)} \nonumber
\end{eqnarray} 
whereas, analogous to the BTK theory\cite{BTK}, the coefficients are complex for $\vert E\vert < \vert \Delta_\pm \vert$ as, 
\begin{eqnarray}
u_{1 \choose 2}=\frac{1}{2} \sqrt{\Big\vert\frac{\Delta_\pm}{E}\Big\vert} e^{i\theta_\pm}\,,\quad \theta_\pm=\tan^{-1}\sqrt{\Big\vert\frac{\Delta_\pm}{E}\Big\vert^2-1}~~~~~~
\label{u_and_v_2}
\end{eqnarray}
and $v_{1 \choose 2}^*=u_{1 \choose 2}$. Here $a,b$ are the complex Andreev reflection amplitudes for the hole with wavevector $K_h=\sqrt{2m(-E+\mu_N)/\hbar^2}$, and the $c,d$ are the normal reflection amplitudes. On the NCS side,  $c_1,\dots,c_4$ are the transmission amplitudes within the NCS in the $\pm$ branches. Here $c=d=0$ as the result of the absence of normal reflection as mentioned before. Hence only $a,b$ are present due to the Andreev mechanism.

\begin{figure}[h]
\includegraphics[scale=0.45]{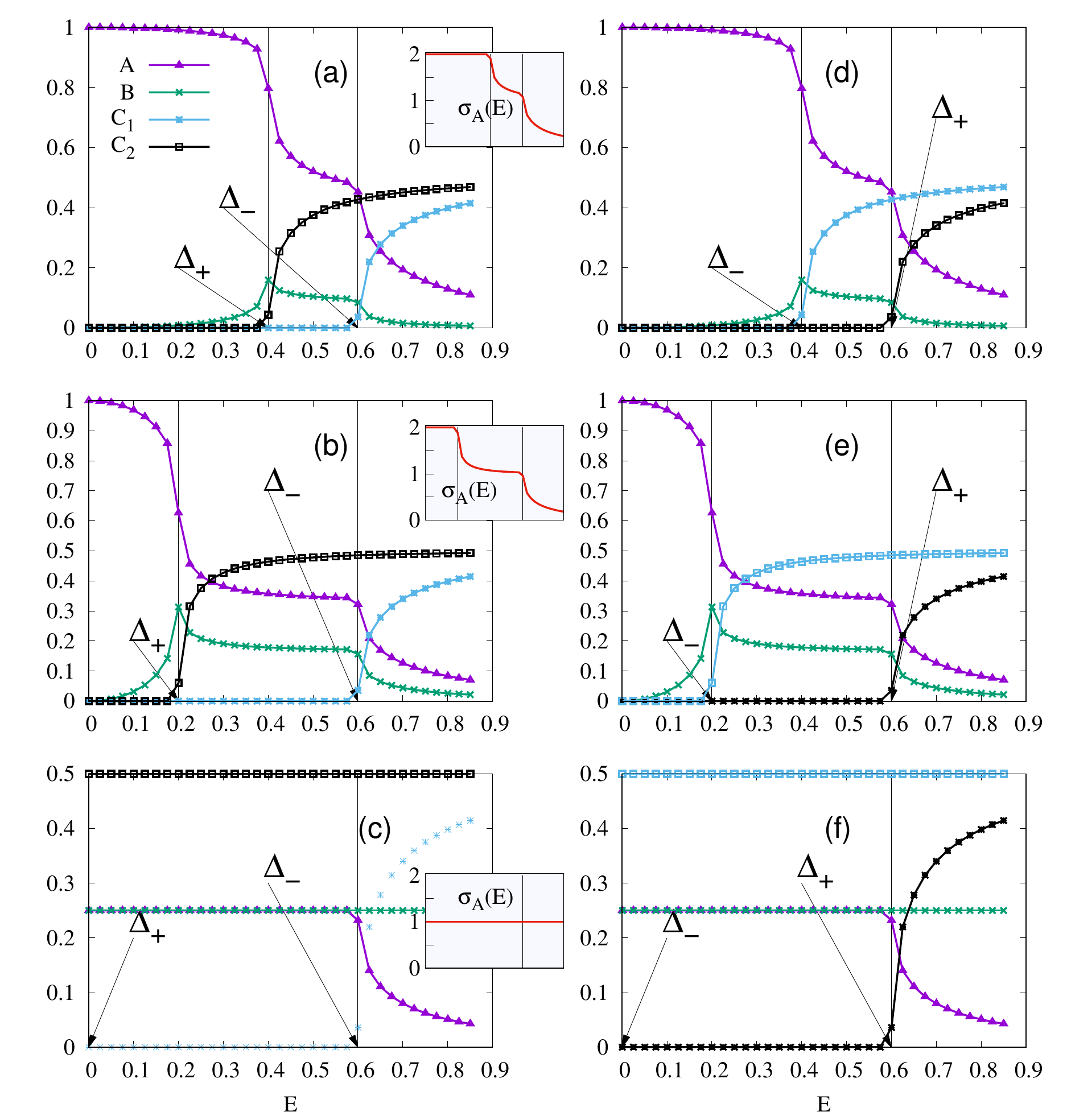}
  \label{FP}\caption{(Color online) Spin resolved Andreev reflection and the transmission probabilities in an N-NCS interface are depicted with Andreev conductance $\sigma_A$ (in units of $2e^2/h$ in the inset for each horizontal case) for three different configurations: a) $\Delta_->\Delta_+>0$, b) same as (a) when $\Delta_+$ is lowered, and c) same as (a) when $\Delta_+=0$. The right columns (c,e,f) correspond to the cases $\Delta_- \leftrightarrow \Delta_+$.}
\label{Andreev_cond}
\end{figure}

The pair potential $\Delta^\lambda_k$ is an isotropic function of $k=\sqrt{k_x^2+k_y^2}$ in the NCS. Due to the homogeneous boundary conditions along the $y$-direction, the $k_y$ is conserved across the boundary. The $\Delta^\lambda_k$ is therefore a function of $E$ of the incident probe particle and its angle of incidence $\phi_i$ on the $N$ side. For normal incidence, we take $\phi_i=0$. This being the case for the perfect isotropy, for weakly anisotropic NCSs the pair potential can be a weak function of the orientaton angle $\phi_0$ of the crystal axes relative to the interface plane in the NCS. In this case, $\Delta^\lambda_k$ can be considered as a fuzzy function of $E$ with a narrow spread given by the degree of anisotropy. In the isotropic limit, fuzziness disappears and $\Delta^\lambda_k=\Delta^\lambda(E)$ becomes  a sharp function of $E$. Assuming this last case and for a given initial energy $E$, there are three different regimes: a) $E < \vert \Delta_- \vert < \vert \Delta_+\vert$ where $\vert u_1^2\vert=\vert v_1^2\vert$ and $\vert u_2^2\vert=\vert v_2^2\vert$, b) $\vert \Delta_- \vert< E < \vert\Delta_+\vert$ where $\vert u_1^2\vert=\vert v_1^2\vert$ and $\vert u_2^2\vert \ne \vert v_2^2\vert$, and c) $\vert\Delta_-\vert< \vert\Delta_+\vert< E$ where $\vert u_1^2\vert \ne \vert v_1^2\vert$ and $\vert u_2^2\vert \ne \vert v_2^2\vert$. We assumed here that $\vert\Delta_-(E)\vert < \vert\Delta_+(E)\vert$ which may or may not be true for all energies [details are in the caption of Fig(\ref{Andreev_cond})]. The results here are unaffected by such details.     

The full solution of the coefficients in Eq.'s(\ref{wf1}), requiring the application of the boundary conditions $\Psi_N(x)=\Psi_S(x)$ and $\Psi_N^\prime(x)=\Psi_S^\prime(x)$ at the $x=0$ interface, has been shown in a large number of works and will not be shown here. The continuity of the current at the interface requires that, 
\begin{eqnarray} 
1-(\vert a\vert^2+\vert b\vert^2)=S_{a,b,c} 
\label{step_like_AC_1}
\end{eqnarray}        
where,
\begin{eqnarray} 
S_a&=&0 , \nonumber \\
S_b&=& \frac{q_2^+}{K_e}\vert c_2\vert^2\,(\vert u_2\vert^2-\vert v_2\vert^2) , \label{step_like_AC_2}\\ 
S_c&=&\frac{q_1^+}{K_e} \vert c_1\vert^2\,(\vert u_1\vert^2-\vert v_1\vert^2) +\frac{q_2^+}{K_e} \vert c_2\vert^2\,(\vert u_2\vert^2-\vert v_2\vert^2) \nonumber   
\end{eqnarray}        
with $S_{a,b,c}$ corresponding to the three cases above. The probabilities $A=|a|^2$ and $B=|b|^2$ are the Andreev reflection probabilities\cite{NCS_review,AR_NCS} for the hole in $x<0$, whereas $C_{1}=\frac{q_1^+}{K_e}|c_1|^2 (\vert u_1\vert^2-\vert v_1\vert^2)$ ,$ C_2= \frac{q_2^+}{K_e} |c_2|^2 (\vert u_2\vert^2-\vert v_2\vert^2)$ are for the transmission probabilities corresponding to the $\pm$ branches. Also, $c_3=c_4=0$ due to the absence of reflection in a semi-infinite geometry in the NCS. Here we will be interested in the double step-like behaviour of the $A=|a|^2$ and $B=|b|^2$ as a result of the Eq.'s(\ref{step_like_AC_2}) as shown in Fig.(\ref{Andreev_cond}).  

Respecting these three energetically different regimes, we examine the AC ($ \sigma_A $) in two distinct cases of the pair potentials: (i) when both $\tilde{\Delta}^\pm$ are nonzero and, (ii) when one of them is zero at a nodal position. The spin dependent Andreev reflection and the transmission coefficients are shown in the Fig.\ref{Andreev_cond}.(a-f) together with the AC $\sigma_A$ calculations in the insets, where the mutual positions of the $\Delta^\pm$ are varied in (a,b,d,e) and one of the pair potentials is assumed to be at the nodal position at $E=0$ in (c,f). Let us concentrate on the inset in Fig.\ref{Andreev_cond}.(a) where the $ \sigma_A $ starts with a plateau at unit conductance corresponding to the configuration $E < \vert \Delta_- \vert < \vert \Delta_+\vert$, i.e. case (a) defined above Eq.(\ref{step_like_AC_1}). If $E$ is between the gaps, as in case (b), the AC moves down to a second plateau. For the case (c), the $ \sigma_A $ gradually disappears as $E$ is increased. If the interval $\vert\Delta^+-\Delta^-\vert$ is changed, the curves change only quantitatively with the double-step behaviour  unchanged [Fig.'s\ref{Andreev_cond}.(b, e)]. We now shift to the second distinct case (ii) above, when one of the pair potentials is zero as shown in Fig.\ref{Andreev_cond}.(c,f) corresponding to a RLN position. The $\sigma_A$ in the inset therein, directly starts at 1/2 at low energies then going through a single plateau before gradually vanishing at high energies. If this RLN is located on the Fermi surface, this should experimentally give rise to a narrow zero bias conductance peak. 

The double steps in the AC is a signature of the three regions with different Andreev reflection properties. We therefore expect that the two distinct steps should always be present where the other details such as the step length and the vertical range of the steps should be material dependent. From the theoretical point of view, double steps clearly point at the weak anisotropy, but in reality there can be a weak dependence on the orientation angle $\phi_0$ in $\Delta^\pm$. This should affect the steepness of the falls in $\sigma_A$ between the plateaus. However, there can be a serious danger on the visibility of the double steps: these characteristic features can be completely erased in the presence of imperfections on the interface as we discuss now.       

We checked the robustness of the double steps against the imperfections on the interface. In this case, we expect that an effective transmission barrier is created on the interface which can easily obscure the ideal AC profile. To understand this effect, we assumed a spin-neutral barrier potential as $Z\delta(x)$ with $Z$ describing the barrier strength and then calculated the AC. Our results confirm the expectation that, the shape of the AC is extremely fragile against the surface imperfections and the double step behaviour is destroyed completely. Highly clean interfaces are therefore needed to observe the ideal double step behaviour. 
 
\begin{figure}
\includegraphics[scale=0.47]{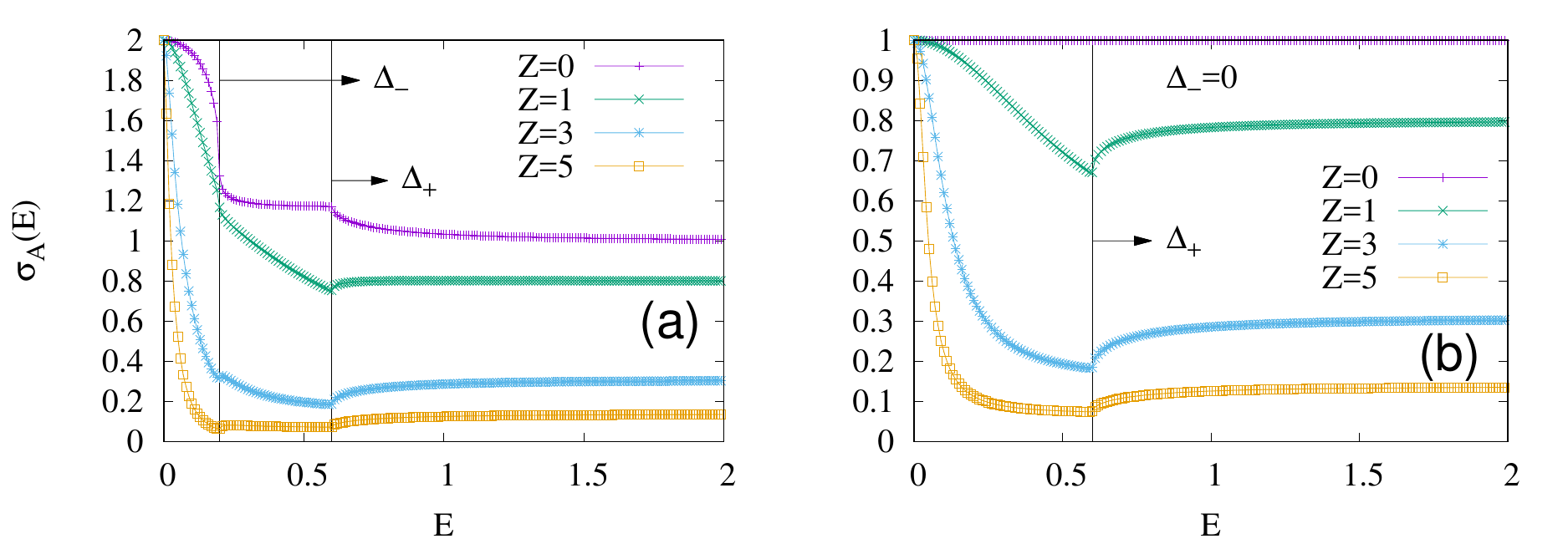}
\label{FP}\caption{(Color online) The $\sigma_A(E)$ (in units of $2e^2/h$) for a semi infinite N-NCS interface including a delta-function-like  interface potential $Z\delta(x)$ when a) both pair potentials are nonzero, b) $\Delta^-=0$ and $\Delta^+ \ne 0$. The $Z$ values are given in units of an energy scale equivalent to $10 meV$.}
\label{Andreev_cond_sp}
\end{figure} 
 
\section{Conclusions}
In this work, we concentrated on the RLNs expected to be present in weakly anisotropic noncentrosymmetric superconductors. Theories incorporating exact momentum dependence of the OPs point at the existence of such nodes. They emerge in the weakly anisotropic limit, in contrast to the ALNs appearing under strongly anisotropic conditions. In an RLN, the pair potential vanishes in at least one reciprocal space point where the singlet and the triplet couplings become locally comparable. In weakly anisotropic systems RLNs are pronounced since their more anisotropic counterparts (the ALNs) are absent. We demonstrated that, they affect the low temperature dynamics that are not explained by the ALNs. In particular, when they are not on the Fermi surface, the RLNs can imitate a BCS like behaviour in the low temperature thermodynamic measurements without giving up the unconventional pairing and the topology. This finding is crucial in that, a number of experimental results on NCSs in favour of {\it trivial s-wave coupling} may need to be reconsidered and these results may happily turn out to be nontrivial. Furthermore several compounds with broken TRS are reported to show a similar behaviour\cite{TRSB_swave}. It is interesting that the picture presented in this work may be also relevant in the time reversal symmetry broken NCSs.  

The topology is classified by the relative position of the RLN in the pair potential with respect to the Fermi wavevector. The latter can be shifted by the chemical potential and the SOC. This brings an additional importance to the RLNs since their topology can be manipulated externally by electrostatic gates and the spin-orbit strength. This can have promise in new research directions based on superconducting node engineering with implications in exotic device applications in topological quantum computing. Finally, the Andreev spectroscopy in N-NCS junctions is an efficient tool for probing the double gap structure of the weakly anisotropic NCSs where RLNs are most likely to be found experimentally.


\begin{thebibliography}{99}
\bibitem{BCS_form} Note that the original spectrum with the full in Eq.(\ref{spectrum}) is not expressible in a generalized BCS form similar to Eq.(\ref{generalized_BCS}).
\end{thebibliography}
\end{document}